\documentclass[12pt]{article}
\usepackage{amsmath}
\usepackage{amsfonts}
\usepackage{amssymb}
\usepackage{hyperref}

\numberwithin{equation}{section}

\newcommand{\ns}{\normalsize} 
\newcommand{\cK}{\mathcal{K}}
\newcommand{\cM}{\mathcal{M}}
\newcommand{\cN}{\mathcal{N}}
\newcommand{\cP}{\mathcal{P}}
\newcommand{\cG}{\mathcal{G}}
\newcommand{\cF}{\mathcal{F}}

\newcommand{\M}{M}

\newcommand{\nn}{\nonumber}

\begin{document}

\begin{titlepage}

\title{
    \vskip 2cm
   {\Large\bf Heterotic type IIA duality with fluxes -- towards the
     complete story}\\[1.5cm]}

\author{{\bf Andrei Micu} \\[1cm]
   {\it\ns Horia Hulubei National Institute of Physics and Nuclear
     Engineering -- IFIN-HH}\\  
   {\ns Atomistilor 407, P.O.~Box MG-6, M\u{a}gurele, 077125, jud
     Ilfov, Romania}\\
   {\tt\ns amicu@theory.nipne.ro}\\[1cm]
}
\date{}

\maketitle

\begin{abstract}
  In this paper we study the heterotic type IIA duality when fluxes
  are turned on. We show that many of the known fluxes are dual to
  each other and claim that certain fluxes on the heterotic side
  require that the type IIA picture is lifted to M or even F-theory
  compactifications with geometric fluxes. 
\end{abstract}

\thispagestyle{empty}

\end{titlepage}

\section{Introduction}
\label{intro}

Flux compactification has been a central research topic in the past
ten years. One of the main interests comes from the fact that fluxes
generate potentials for certain moduli fields which can be fixed in
this way. Type II theories have been studied extensively and there are
examples when all moduli have been fixed by fluxes. This is however a
rather ungeneric situation and most of the time there are moduli
fields which do not appear in the flux generated superpotential. The
number and types of moduli fields which appear/do not appear in the
superpotential largely depends on the scheme of compactification
considered. It is therefore important to know all the possible fluxes
which can be turned on in a given setup.
A powerful way to control the fluxes is by using
dualities. There exist several instances when new fluxes have been
discovered in this way. It was first realised in \cite{DRS} that T and
S-dualities trade flux type IIB backgrounds for torsional heterotic
geometries.\footnote{For recent similar examples where non-geometric
  heterotic solutions are dualised to geometric F-theory backgrounds
  see \cite{BS,OMS}.}
Other examples include mirror symmetry where the NS-NS
flux is mapped to compactifications on manifolds with $SU(3)$
structure \cite{GLMW,GM}. The same was noticed in toroidal
compactifications where again the NS-NS flux was mapped by T-duality
into compactifications on twisted tori \cite{KST}. More complete
pictures of these dualities in the presence of fluxes have appeared
once manifolds with $SU(3) \times SU(3)$ structure 
\cite{GMPT1}-\cite{GSN}
T-folds
\cite{CH1}-\cite{HRE2}
or the notion of non-geometric
\cite{STW} fluxes have been introduced. A systematic way of finding
all possible fluxes in a given setup is to start from a certain background
with fluxes and apply all possible duality transformations (like S and
T-dualities, $SL(2,\mathbf{Z}$) or even exceptional duality groups)
until no new fluxes are generated. In this way, many new fluxes have
been discovered in  
\cite{ACFI}-\cite{AACG}.

One very interesting case, which we will study in this paper, is the
heterotic type IIA duality with fluxes. The interest comes mainly from
the fact that the two compactifications schemes are rather different
in nature. This duality in the presence of fluxes was first analysed
in \cite{CKKL} where only a small part of the fluxes which can be
turned on in the heterotic picture were mapped the type IIA side. This
was further extended in \cite{LM3} where 
it was realised that fluxes on the heterotic side correspond to
geometric fluxes (manifolds with $SU(3)$ structure) on the type IIA
side. Finally, in \cite{ABLM} it was argued that certain fluxes in the
heterotic picture do not find a correspondent in type IIA
compactifications, but one has to lift this to M-theory
compactifications on seven-dimensional manifolds with $SU(3)$
structure. 

In this paper we continue the study of heterotic type IIA duality with
fluxes. We first give an extended version of the results in
\cite{ABLM}. Considering fluxes which gauge isometries in the vector
multiplet sector, we show that the whole amount of fluxes which appear
in the M-theory picture can be understood on the other side by
heterotic compactifications with duality twists as spelled out in
\cite{DH,RES}. We further extend this analysis and conjecture that
``R-fluxes'' in the heterotic picture correspond to F-theory
compactifications on eight-dimensional manifolds with $SU(3)$
structure.

We make a similar analysis for the fluxes which gauge isometries in
the hyper-multiplet sector. Even though in this case there are more
fluxes on the type IIA side which gauge isometries in the
hyper-multiplet sector, we still find that for certain setups on the
heterotic side one needs to consider M and even F-theory
compactifications.\footnote{M-theory compactifications on twisted
  seven-dimensional manifolds of the type we consider here were
  recently studied in \cite{LPV}.} 
Therefore the broadest picture we shall obtain will
be that F-theory compactifications on eight-manifolds which are
obtained by fibering a Calabi--Yau over a torus in a particular way
are dual to heterotic string compactifications on six manifolds with
SU(2) structure. However even in this case there will still
be fluxes on either side which are unaccounted for. It is possible
that most of this mismatch comes from the fact that on the heterotic
side we ignore a large part of the hyper multiplet space, namely that
related to the gauge bundle deformations as this space can not be
treated in a generic fashion. Hopefully for a well defined setup one
can account for all the fluxes which can be turned on.

The outline of the paper is the following. In section \ref{bg},  we
start by recalling few facts about compactifications of type IIA and
heterotic strings to four dimensions with $N=2$ supersymmetries. We
do this mostly for fixing the conventions and notations. We also
explain briefly the effect of turning on fluxes in order to
motivate our analysis. In section \ref{vmg}, we analyse fluxes on both
heterotic and type IIA side which gauge isometries in the vector
multiplet sector. We first recall M-theory compactifications on seven
dimensional manifolds with $SU(3)$ structure which lead to gaugings in
the vector multiplet sector. Then we show that all these fluxes can be
found in heterotic compactifications with duality twists. Then we
conjecture that R-fluxes in heterotic compactifications can be
accounted for in a F-theory setup. We also make some speculations
about how one can obtain a charged dialton in the heterotic
picture. In section \ref{hmg} we turn to those fluxes which gauge
isometries in the hyper-multiplet sector. After we review the known
facts about the duality when ordinary fluxes are turned on we recall
the results of heterotic compactifications on manifolds with $SU(2)$
structure and show that part of the content of the theory can 
again be obtained on the type IIA side if one considers M-theory
compactifications on manifolds with $SU(3)$ structure. To recover the
full picture we conjecture again that one needs to go to F-theory
compactifications. In the appendices we gathered the most important
information about $N=2$ gauged supergravity in four dimensions and about
the structure of the vector multiplet sector in heterotic string
compactifications. We also derive here the gaugings which appear in
the vector multiplet sector in the compactifications with duality
twists.

\section{Preliminaries}
\label{bg}

As explained in the Introduction, we are interested in the duality
between heterotic and type IIA string compactified to four dimensions
with $N=2$ supersymmetry. On the {he\-te\-ro\-tic} side such situations can be
obtained upon compactifications on $K3 \times T^2$ manifolds, while in
type IIA we deal with $K3$ fibered Calabi--Yau
manifolds \cite{AL}. $N=2$ supersymmetry ensures that the scalar sector of the
theory splits into a part describing the vector multiplets and one
which describes the hyper-multiplets. In the presence of fluxes
certain isometries of the scalar manifold are gauged and this leads to
a mixing between these two sectors. However it still makes sense to
distinguish between them and moreover to distinguish between fluxes
which gauge isometries in the vector and those which gauge isometries
in the hyper-multiplet sector. In the following we shall review these
compactifications without fluxes and in the end explain briefly the
effect of turning on fluxes.

\subsection{Heterotic on $K3 \times T^2$}
\label{hk3t2}

Let us start with the heterotic side. As explained above we consider
the heterotic string compactified on $K3 \times T^2$.  For simplicity
we shall not take into account the non-Abelian structure of the
heterotic supergravity in ten dimensions and restrict ourselves to the
Cartan algebra of the ten-dimensional gauge group.  As the aim of the
paper is to make general remarks about the heterotic--type IIA duality
with fluxes, these aspects are not going to be important for us. We
nevertheless have to keep in mind that we have to satisfy the
heterotic Bianchi identity and turn on a certain gauge bundle on $K3$
which actually breaks the original gauge group.  Therefore the precise
resulting gauge group in four dimensions is not known unless one
specifies explicitly the gauge bundle which is turned on in the
background and we shall work with a generic $U(1)^{n_v+1}$
for some number $n_v$ of vector
multiplets which depends on the details of the compactification
and which we shall leave arbitrary. Part of the gauge fields come from
the Cartan sub-algebra of the surviving gauge group while four of the
gauge fields emerge from the Kaluza-Klein vectors on the torus as well
as form the $B$-fields with one leg on the torus.\footnote{In general
  it may happen that some of these gauge fields are also broken in
  the compactification due to the particular gauge bundle which is
  chosen on $K3$. These gauge fields are particularly important in the
  gaugings which occur in the vector multiplet sector and we shall
  assume that they survive the compactification to four dimensions. In
  the absence of (some of) these gauge fields, many -- and probably,
  the most interesting -- of the vector multiplet gaugings will be
  absent.} One of these gauge fields (or a combination thereof) is the
graviphoton and is part of the gravity multiplet, while the remaining
gauge fields are part of vector multiplets. In four dimensions, each
vector multiplet also contains one complex scalar, and in the present
setup $2 \cdot (n_v-3)$ of them come from the ten-dimensional gauge
fields on the torus, four from the $T^2$ moduli and from the $B$-field
with both legs on the torus, and the last two from the ten-dimensional
dilaton and from the universal axion, the Poincare dual of the
four-dimensional $B$-field. All these fields span the homogeneous
space
\begin{equation}
  \label{vms}
  \cM_v = \frac{SU(1,1)}{U(1)} \times \frac{SO(2,n_v-1)}{SO(2) \times
    SO(n_v-1)} \; .
\end{equation}

Beside the vector multiplets there is a certain number of
hyper-multiplets. Among the scalars in the hyper-multiplets there are
the $K3$ moduli together with the $B$-field on $K3$ and the $K3$
volume which span the homogeneous space
\begin{equation}
  \label{hms}
  \cM_h = \frac{SO(4,20)}{SO(4) \times SO(20)}
\end{equation}
In the hyper-multiplets there are also the scalar fields which
parameterise the deformations of the gauge bundle. The number of these
fields and the geometry of the full space spanned by these scalars
highly depends on the background gauge bundle and therefore in the
following we shall only concentrate on the part which which is due to
the $K3$ moduli. 

The four-dimensional theory obtained from this compactification is a
$N=2$ supergravity coupled to the vector multiplets and
hyper-multiplets described above. The bosonic part of the action is
given by the general formula given in the appendix, \eqref{s4het}.  In
this case $x^i=(s,t,u,n^a), i=1, \ldots, n_v; ~ a = 4, \ldots, n_v$
denote collectively 
the complex scalars in the vector multiplets, which in terms of the
compactification fields are given in \eqref{complex} and $F^I=d A^I$
denote the Abelian field strengths of the vector fields
$A^I,~ I=0, \ldots, n_v$. Moreover, $g_{i \bar j}$ is the corresponding K\"ahler metric
derived from the K\"ahler potential 
\begin{equation}
  \label{Kh}
  K= -\ln{i(\bar s-s)} - \ln{\tfrac14\big[(t- \bar t)(u - \bar u) -
    (n^a - \bar n^a)(n^a - \bar n^a) \big]} \; .
\end{equation}
and the gauge coupling matrix is given by \eqref{Nhet}. All the
information related to the vector multiplet sector is encoded in
\eqref{XFhet}, but due to the absence of the prepotential the
derivation of the gauge coupling matrix has to follow the general line
developed in \cite{N2rev} rather than using the formula \eqref{NN2}

Finally, the scalars in the hyper-multiplets $q^u$ and the
corresponding quaternionic metric $h^{uv}$ can not be precisely
written down in general and therefore we shall mostly concentrate on
the scalars coming from the $K3$ moduli. Let us briefly recall how
these fields appear. $K3$ is a hyper-K\"ahler manifold and thus has a
triplet of complex structures $J^x, ~ x =1,2,3$. The metric
deformations on $K3$ can be parameterised by deformations of these
complex structures. Let us expand the complex structures $J^x$ in a
basis of harmonic forms on $K3$, $\omega^A, ~ A= 1, \ldots , 22$
\begin{equation}
  \label{Jexph}
  J^x = \zeta^x_A \omega^A \; .
\end{equation}
The parameters of these expansions, $\zeta^x_A$, will appear in the
effective four-dimensional theory as scalar fields. However, not all
of them represent independent degrees of freedom but their variations
are subject to the constraints
\begin{equation}
  \label{zetacons}
  \zeta^x_A \delta \zeta^{yA} = 0 \; ,
\end{equation}
which leave us with 57 possible deformations. To these we add the
volume modulus as well as the massless modes which come from the
$B$-field which we parameterise as
\begin{equation}
  \label{bexp}
  B = b_A \omega^A \; .
\end{equation}
Altogether these fields span the homogeneous space $SO(4,20)/SO(4)
\times SO(20)$.

One important thing to keep in mind from this section is that for the
heterotic compactification the vector multiplet sector is governed
entirely by the $T^2$ part of the compactification while the
hyper-multiplet sector comes from the $K3$ part. Therefore, in this
case, the split between vector and hyper-multiplets comes in naturally
due to the split of the compactification manifold. We shall see
shortly that the same sort of split can be observed also for the
fluxes, as fluxes in the $T^2$ have as effect gaugings of the
vector-multiplet isometries, while $K3$ fluxes gauge isometries in the
hyper-multiplet sector.

\subsection{Type IIA on Calabi--Yau three-folds}
\label{iiacy}

Let us now discuss the compactification of type IIA supergravity on
Calabi--Yau manifolds. For the moment we shall present the general
features, and later we shall specialise to the case of $K3$ fibered
manifolds which are relevant in the heterotic-- type IIA duality.
We keep the discussion short and for more details we refer the reader
to \cite{LM2,Sebth} which we closely follow. In
type IIA compactifications on Calabi--Yau manifolds there is also a
natural splitting between the vector and hyper-multiplets. This
however comes from the Calabi--Yau moduli space property that it
splits into a product of the space of K\"ahler class deformations and
the space of complex structure deformations. The first gives the
scalars in the vector multiplets while the latter together with the RR
axions as well as the dilaton and four-dimensional axion give the
scalars in the hyper-multiplets. Altogether the effective theory in
four dimensions is $N=2$ supergravity coupled to $n_v= h^{1,1}$ vector
multiplets and $n_h=h^{2,1}+1$ hyper-multiplets, where $h^{1,1}$ and
$h^{2,1}$ denote the dimensions of the corresponding cohomology
groups of the Calabi--Yau manifold under consideration.

Let us denote by $\omega_i$ and $\tilde \omega^i, ~ i =1, \ldots ,
h^{1,1}$ the harmonic $(1,1)$ and $(2,2)$ forms and by $(\alpha_A,
\beta^A), ~ A = 0,1,\ldots, h^{2,1}$ a real basis for the harmonic
3-forms on the Calabi--Yau manifold.  These forms are taken to satisfy
\begin{equation}
  \label{norm}
  \int_X \omega_i \wedge \tilde \omega^j = \delta_i^j \; , \qquad \int_X
  \alpha_A \wedge \beta^B = \delta_A^B \; ,
\end{equation}
with all other integrals vanishing. 

In order to obtain the low energy degrees of freedom we expand the
ten-dimensional form fields in the above harmonic forms. From the
expansion of the 3-form potential $C_3$ we obtain $h^{1,1}$ gauge
fields, $A^i$, and $2(h^{2,1}+1)$ scalars (RR-axions), $\xi^A$ and
$\tilde \xi_A$
\begin{equation}
  \label{c3exp}
  C_3 = A^i \omega_i + \xi^A \alpha_A - \tilde \xi_A \beta^A \; .
\end{equation}
The remaining fields come from the metric deformations of the
Calabi-Yau manifold. Let us denote by $x^i$ the complexified K\"ahler
deformations which are given by
\begin{equation}
  \label{Kdef}
  J + iB = x^i \omega_i  \; ,
\end{equation}
where $J$ denotes the K\"ahler form on the Calabi--Yau manifold
and $B$ is the $B$-field on the internal space. We can introduce
projective coordinates in the form
\begin{equation}
  \label{xX}
  x^i = \frac{X^i}{X^0}\; ,
\end{equation}
and it turns out that the special K\"ahler geometry on this space is
given by the prepotential
\begin{equation}
  \label{prepot}
  \cF = - \frac16 \frac{\cK_{ijk} X^i X^j X^k}{X^0} \; ,
\end{equation}
where $\cK_{ijk}$ are the triple intersection numbers on the
Calabi--Yau manifold 
\begin{equation}
  \label{int}
  \cK_{ijk}= \int_{Y_6} \omega_i \wedge \omega_j \wedge \omega_k \; .
\end{equation}
Finally, for the vector multiplet sector, the gauge coupling matrix
$\cN_{IJ}$ is given by the general $N=2$ formula \eqref{NN2} for the
prepotential \eqref{prepot}.

The complex structure deformations can be obtained from the expansion
of the holomorphic $(3,0)$ form on the Calabi--Yau manifold in the
(real) basis of three-forms $(\alpha_A , \beta^A)$
\begin{equation}
  \label{Oexp}
  \Omega = Z^A \alpha_A - \cG_A \beta^A \; ,
\end{equation}
where $\cG_A$ are the derivatives of the prepotential corresponding to
the special K\"ahler geometry which describes the complex structure
moduli space and $Z^A$ are projective coordinates on this space.

With the above notations the action for the effective theory obtained
by compactifying the type IIA string on Calabi--Yau manifolds can be
put in the form \eqref{s4het} where the hyper-scalars $q^u$ denote
collectively $q^u= (z^a, \phi, a, \xi^A, \tilde \xi_A)$, with $z^a$
the complex structure deformations given by
\begin{equation}
  z^a = \frac{Z^a}{Z^0} \; , \quad a = 1, \ldots, h^{2,1} \; ,
\end{equation}
$\phi$ the dilaton, $a$ the axion which is Poincare dual to the
four-dimensional $B$-field, and $\xi^A$ and $\tilde \xi_A$ defined in
\eqref{c3exp}.

\subsection{Turning on fluxes}

Let us explain Briefly the effect of turning on fluxes in the above
compactifications. 
We first concentrate on the fluxes which can be turned on in the
heterotic picture. The p-forms available for turning on fluxes are the
gauge field strengths two forms $F^I$ and the three-form $H$, the
field strength of the NS-NS $B$-field. Inside
$K3$ there are 22 two-cycles on which we can turn on the fluxes $F^I$
and these lead to gaugings in the hyper-multiplet sector. There is also
the possibility that we turn on the the fluxes $F^I$ along $T^2$ and
this turns out to gauge isometries in the vector multiplet
sector. Finally $H$ can only be turned on with one leg on the torus
and two legs along some $K3$ two-cycle. Since the $B$-field with one
leg on the torus is again one of the four-dimensional gauge fields we
can think of the $H$ fluxes 
again as fluxes for these gauge fields on $K3$. Therefore, we shall only
distinguish between fluxes strictly inside $K3$ and fluxes along
$T^2$. The first gauge isometries in the hyper-multiplet space while
the latter gauge isometries in the vector multiplet space.

Other types of fluxes on the heterotic side can be obtained by
deforming the {com\-pac\-ti\-fi\-cation} manifold. We can consider
compactifications with duality twists -- also known as T-folds. Such
compactifications, when applied to our case lead to gaugings in the
vector multiplet sector as we shall explain in section
\ref{vmg}. Other deformations include manifolds with $SU(2)$ structure
and it turns out that these lead to gaugings in the hyper-multiplet
sector. 

On the type IIA side there are several fluxes available. First of all
the is the three-form flux for $H$, the field strength of the NS-NS
antisymmetric tensor $B$. Moreover, we can turn on RR fluxes which
comprise all even forms. All these fluxes gauge isometries in the
hyper-multiplet sector
There are also generalisations of these fluxes which include
geometric fluxes (manifolds with $SU(3)$ structure) or non-geometric
fluxes (manifolds with $SU(3) \times SU(3)$ structure). We shall
discuss all these fluxes in more detail in section \ref{hmg}, but for
the moment it is important to note that all these fluxes only gauge
isometries in the hyper-multiplet sector.

The purpose of the rest of the paper is to try to match the various
fluxes discussed above between the heterotic and type IIA picture. As
we have already pointed out there are no fluxes strictly within type
IIA theory which can lead to gaugings in the vector multiplet sector.
It was proposed that the heterotic fluxes which
gauge isometries in the vector multiplet space can actually be
described in M-theory rather than in type IIA. We shall review this
proposal in the next section and present arguments that further
extensions of this proposal involve also F-theory.

\section{Vector multiplet gaugings}
\label{vmg}

From the brief review in the previous section we have seen that in the
case of the heterotic string compactifications we can easily obtain
gaugings in both hyper- and vector multiplet sector. On the other
hand, in the type IIA picture there are no gaugings in the vector
multiplet sector. In this section we shall concentrate on the vector
multiplet sector and explain how we can obtain gaugings in this sector
in the context of type IIA theories. 

\subsection{Heterotic compactifications with gauge field fluxes on $T^2$}

At the beginning of this sections let us briefly recall the effect of
fluxes which we can turn on on $T^2$ in heterotic string
compactifications. 
Let us consider fluxes of the type
\begin{equation}
  \label{T2flux}
  \int_{T^2} F^a = f^a  \; , \quad a=4,\ldots, n_v \; .
\end{equation}
Such compactifications were considered in  \cite{KM,LM1} and here we
only briefly recall the results. Later on we will also present a more
detailed calculation from where these results can be obtained. We will
only be interested in the vector multiplet sector whose structure is
explained in appendix \ref{appB}.

The result is that some of the scalars in the vector multiplets become
charged and their covariant derivatives read
\begin{equation}
  \label{vmfl}
  \begin{aligned}
    D t = & ~ \partial t - \sqrt 2\, n^a f^a A^1 + f^a A^a \; ,\\
    D n^a = &  ~ \partial n^a - \tfrac{1}{\sqrt 2}\, f^a (A^0 + u A^1) \; ,
  \end{aligned}
\end{equation}
Moreover the gauge group becomes non-Abelian and the field strengths
for the gauge fields are given by
\begin{eqnarray}
  \label{hetfs}
  F^0 & = & d A^0 \; , \nn \\
  F^1 & = & d A^1 \; , \nn \\
  F^2 & = & d A^2 +  f^a A^a \wedge A^1 \; , \\
  F^3 & = & d A^3 -  f^a A^a \wedge A^0 \; ,\nn  \\
  F^a & = & d A^a -  f^a A^0 \wedge A^1 \; . \nn
\end{eqnarray}

There is also a potential which is generated, but it is completely
fixed by the $N=2$ supersymmetry from the above data and hence we
shall not be concerned with it in the following.

\subsection{M-theory compactifications on seven-dimensional 
manifolds  with $SU(3)$ structure} 
\label{MS3}

We shall now explain what is th correspondent in the type IIA setup of
the  picture presented above. As it was shown in \cite{ABLM}, we are
led to consider M-theory compactifications on seven-dimensional
manifolds. We review the results in \cite{ABLM,LPV} below.
The main insight for the origin of the fluxes which produce gaugings
in the vector multiplet sector comes from studying the heterotic type
IIA duality with heterotic fluxes on $T^2$ from the perspective of the
five-dimensional duality between heterotic string compactified on $K3
\times S^1$ and M-theory compactified on Calabi--Yau manifolds. Upon
further compactifying on a circle we end up with
the heterotic type IIA duality in four dimensions that we are
interested in. From this point of
view, the heterotic fluxes appear only in this last step. These fluxes
can be thought of as monodromies of the scalars in the 5d vector
multiplets around the circle which takes us down to four dimensions. 
We can try to do something similar in the M-theory case in the $S^1$
compactification. The 5d vector multiplet scalars come from the
K\"ahler moduli of the Calabi--Yau manifold and the monodromies around
the circle imply that actually the Calabi--Yau manifold is fibered
over the circle. Denoting again the harmonic two-forms on the
Calabi--Yau by $\omega_i$ we describe the monodromy by
\begin{equation}
  \label{do}
  d \omega_i = \M_i^j \omega_j \wedge dz
\end{equation}
where the constants $\M_i^j$ form the twist (monodromy) matrix while $dz$
describes the circle direction.\footnote{Note that we are indeed
  dealing with a $SU(3)$ structure as manifolds with $SU(3)$ holonomy
  in seven dimensions necessarily have $M_i^j=0$ in the equation
  above.} The compactification on such manifolds was proposed and
carried out in \cite{ABLM} and in the following we briefly summarise the
results. The action in four dimensions is again given by \eqref{s4het} but
now with covariant derivatives replacing the ordinary derivatives on
the scalar fields in the vector multiplets
\begin{equation}
  \label{kvA}
  Dx^i = dx^i - k^i_I A^I \ , \qquad\mathrm{with}\qquad  k_0^j = - x^k \M_k^j
  \; , \qquad k_i^j = \M_i^j \; .
\end{equation}
Moreover, the field strengths of the four-dimensional gauge fields are
modified as
\begin{equation}
  \label{strcon}
  F^I = d A^I + \tfrac12 f^I_{JK}A^J\wedge A^K \ , \qquad\mathrm{with}\qquad
  f_{ij}^0 = 0 = f_{ij}^k  \; , \qquad f_{i0}^j = - \M_i^j  \; ,
\end{equation}
where $i,j = 1,\ldots, n_v$ and $I,J = 0,1,\ldots, n_v$.
Finally the action has to be supplemented by the Chern-Simons
generalised term
\begin{equation}
  \label{gCS}
  S_{gCS} = - \tfrac16 \int_{M_4} \M_i^l \cK_{jkl} A^i \wedge A^j \wedge
  dA^k \; , \nn 
\end{equation}
which appears in addition to the standard action \eqref{s4het} because of
the lack of invariance of the prepotential under the gauge
transformations \cite{ABLM}. 

In the above setup, the parameters $\M_i^j$ are subject to the
constraint 
\begin{equation}
  \label{const}
  \M_i^l \cK_{jkl} + \M_j^l \cK_{kil} + \M_k^l \cK_{ijl} =0 \; ,
\end{equation}
which comes from the fact that the volume of the Calabi--Yau manifold
(which in five dimensions is a member of a hyper-multiplet) should not
change as we move along the circle. 

So far the discussion was general and can apply in principle for any
Calabi--Yau manifold. The constraint \eqref{const} on the other hand
tells us that the moduli space of K\"ahler deformations admits an
isometry which is not a generic property of Calabi--Yau manifolds. 
For the heterotic--type IIA duality, the relevant Calabi--Yau
manifolds are $K3$ fibrations over a $\mathbf{P_1}$ base \cite{AL}
and the intersection numbers have the following structure
\begin{equation}
  \label{intno}
  \cK_{123} = -1 \; , \qquad \cK_{1ab} = 2 \delta_{ab} \; , \qquad
  a,b=4, \ldots , h^{(1,1)} = n_v \; ,
\end{equation}
where the index 1 denotes the base and the indices 2 and 3 denote
other two-cycles which are singled out. In such a case the solution to
the constraint \eqref{const} can be parameterised as
\begin{equation}
  \label{indpar}
  m_2   \equiv \M_2^2  \; , \qquad m_a \equiv  \M_a^2 \; , \qquad
  m_3  \equiv \M_3^3 \; , \qquad \tilde m_a \equiv \M_a^3 \; , \qquad
  m_b^a \equiv -  \M_a^b  \  ,
\end{equation}
where $m^a_{b} = -m^b_{a}$ and the other matrix elements are then given by
\begin{equation}
  \label{deppar}
  \begin{aligned}
    \M_2^a  =  \tfrac12 \tilde m_a  \; , & \qquad \M_3^a = \tfrac12 m_a \; ,
    \qquad  \M_a^a = - \tfrac12 M_1^1\ =\ \tfrac12 (m_2 +  m_3) \;
    ,\\
    & M_1^{2,3} = M_1^a =  M_a^{1} = M_{2,3}^1  = M_2^3= M_3^2 = 0
\ .
  \end{aligned}
\end{equation}

One of the main task of this section is to find the heterotic
correspondent of all the parameters above. In \cite{ABLM} it was shown
that the parameters $\tilde m_a$ correspond to gauge field fluxes on
the heterotic side. The case $m_2+m_3 \ne 0$ is a bit more subtle and
we shall discuss some ideas at the end of this section. So, for the
moment 
we consider that $m_2 + m_3 = 0 $ and show that all the parameters
above can be recovered in the compactification of the heterotic
supergravity with duality twists.

In order to be able to compare the type IIA (M-theory) picture with
the heterotic one we should first perform an electric-magnetic duality
in order to be in the same symplectic frame on both sides. This
requires to exchange one of the gauge fields ($A_1$ in the case at
hand) with its magnetic dual. For the case $m_2 = - m_3= m$ one can
immediately see from \eqref{kvA} that no scalar fields are charged
under $A^1$ and from \eqref{strcon} that the field strength for this
vector field is simply $F^1 = d A^1$. The only place where this gauge
field appears non-trivially is in the generalised Chern-Simons term
\eqref{gCS}. Let us write this term in more detail for the specific
parameters from \eqref{indpar}.
\begin{eqnarray}
  \label{CSint}
  S_{gCS} & = & \tfrac13 \int \left( m A^2 \wedge A^3 - \tilde m_a A^2 \wedge
    A^a - m_a A^3 \wedge A^a - m_b{}^a A^b \wedge A^a \right) \wedge
    d A^1 \\
    & & - \tfrac16 \int d\left( m A^2 \wedge A^3 - \tilde m_a A^2 \wedge
    A^a - m_a A^3 \wedge A^a - m_b{}^a A^b \wedge A^a \right) \wedge A^1
  \nn
\end{eqnarray}
Integrating by parts in the last term we end up with 
\begin{equation}
  \label{CSf}
  S_{gCS} = \tfrac12 \int \left( m A^2 \wedge A^3 - \tilde m_a A^2 \wedge
    A^a - m_a A^3 \wedge A^a - m_b{}^a A^b \wedge A^a \right) \wedge
    d A^1 \; .
\end{equation}
We see that the field $A^1$ actually appears in the action
only through its field strength $F^1= d A^1$ and hence, can be easily
dualised. The result of the dualisation is that the generalised
Chern-Simons term disappears while the field strength of the magnetic
dual gauge field $\tilde A^1$ has the form
\begin{equation}
  \label{G}
  G^1 = d \tilde A^1 - m A^2 \wedge A^3 + \tilde m_a A^2 \wedge A^a +
  m_a A^3 \wedge A^a + m_b{}^a A^b \wedge A^a \; .
\end{equation}
The remaining field strengths have the form
\begin{eqnarray}
  \label{fsiia}
  F^0 & = & dA^0 \; ; \nn \\
  F^2 & = & d A^2 + m A^0 \wedge A^2 + m_a A^0 \wedge A^a \; \\
  F^3 & = & d A^3 - m A^0 \wedge A^3 + \tilde m_a A^0 \wedge A^a \;
  \nn \\
  F^a & = & d A^a + \tfrac12 \tilde m_a A^0 \wedge A^2 + \tfrac12 m_a
  A^0 \wedge A^3 + m_a^b A^0 \wedge A^b \; . \nn 
\end{eqnarray}
We can therefore read off the following non-vanishing structure
constants of the gauge algebra 
\begin{equation}
  \label{sciia}
  \begin{aligned}
    f_{23}^1 = - m \; & ; \quad f_{2a}^1 = \tilde m_a \; ; \quad f_{3a}^1
    = m_a \; ; \quad f_{ab}^1 = 2 m_a{}^b \; ; \quad f_{02}^2 = m \; ;
    \quad f_{0a}^2 = m_a \; ;\\ 
    f_{03}^3 = -m \; & ; \quad f_{0a}^3 = \tilde m_a \; ; \quad
    f_{02}^a = \tfrac12 \tilde m_a \; ; \quad f_{03}^a = \tfrac12 m_a
    \; ; \quad f_{0a}^b = m_a{}^b \; .     
  \end{aligned}
\end{equation}
Finally the gauged isometries are the same as in
\eqref{kvA} and we can write explicitly
\begin{eqnarray}
  D_\mu x^1 & = & \partial_\mu x^1  \nn \\
  \label{cdiia}
  D_\mu x^2 & = & \partial_\mu x^2 + (m x^2 + m_a x^a) A^0_\mu -m
  A^2_\mu - m_a A^a_\mu \; ;  \\
  D_\mu x^3 & = & \partial_\mu x^3 + (-m x^3 + \tilde m_a) A^0_\mu +
  mA^3_\mu - \tilde m_a A^a \; ; \nn \\
  D_\mu x^a & = & \partial_\mu x^a + (\tfrac12 \tilde m_a x^2 +
  \tfrac12 x^3 - m_a^b x^b) A^0_\mu - \tfrac12 \tilde m_a A^2_\mu - \tfrac12
  m_a A^3_\mu + m_a^b A^b_\mu \; . \nn
\end{eqnarray}
Reading off the killing vectors from the above
one can check explicitly that the relation
\begin{equation}
  \label{comm}
  [k_I , k_J] = f_{IJ}^K k_K \; ,
\end{equation}
holds for the structure constants \eqref{sciia}.

\subsection{Heterotic compactifications on $K3 \times T^2$ with
  duality twists}

In the previous subsection we have reviewed the structure of M-theory
compactifications on manifolds with $SU(3)$ structure which produce
gaugings in 
the vector multiplet sector. In the following we will show that the
same result can be obtained from heterotic string compactifications
with duality twists. The $K3$ part of the compactification will be a
``spectator'' in the current section and we shall be interested only in
the $T^2$ part. The compactification on $K3$ will be assumed to follow
in a straightforward manner. 

Heterotic compactifications on tori in the presence of fluxes was
initially studied in \cite{KM}. Here it was shown that the gauge field
fluxes, the $H$-fluxes and the geometric fluxes coming from the
twisting of the compactification torus fit nicely in the $O(d,d+16)$
framework of the compactified theory. More recently this setup was
generalised in order to include compactifications with (T-)duality
twists which are also termed as non-geometric backgrounds
\cite{DH,RES}. In this section we shall use these recent results in
order to show that one can obtain precisely all the flux parameters
which were described in the previous section. 

The main idea of the duality twists compactification is that one can
split the $d$-dimensional torus into a product $T^{d-1} \times
S^1$. The compactification on the $d-1$ torus gives a theory with a
$O(d-1,d+15)$ duality symmetry. This can be further compactified on
the last $S^1$ allowing also the fields to vary according to the
$O(d-1,d+15)$ duality symmetry. In the case at hand we are interested in
$T^2$ compactifications and therefore we split it as $S^1 \times S^1$
and perform a duality twist compactification on the second $S^1$. We
have to keep in mind that the $K3$ part of the compactification
generically breaks also some of the Cartan generators of the original
gauge group and therefore the duality group may not be the full
$O(1,17)$ group and we shall generically denote it by $O(1,n_v-2)$,
where $n_v$ denotes the number of vector multiplets in the final
four-dimensional theory. 

The most general twist matrix as spelled out in \cite{DH,RES} has the
form
\begin{equation}
  \label{twist1}
  N_N{}^P = \left(
  \begin{array}{ccc}
    f & 0 & M^b \\
    0 & -f & W^b \\
    -W_a & -M_a & S_a{}^b \\
  \end{array}
  \right) \; .
\end{equation}
Based on duality arguments, the structure constants of the gauged $N=4$
supergravity were found to be given in terms of the twist matrix as
\begin{equation}
  \label{schet}
  f_{0N}^P = N_N{}^P \; \quad f_{NP}^1 = N_{NP} \; ,~ N,P = 2,3,\ldots , n_v
\end{equation}
where the indices 0 and 1 denote the directions in the gauge field
space given by the Kaluza-Klein vector on $S^1$ and the $B$-field with
one leg on $S^1$ respectively. The indices of the twist matrix $N$ are
raised and lowered with the $O(1,n_v-2)$ invariant 
\begin{equation}
  \label{L}
  L = \left(
    \begin{array}{ccc}
      0 & 1 & 0 \\ 
      1 & 0 & 0 \\
      0 & 0 & \mathbf{1}_{n_v-3}
    \end{array}
    \right) \; .
\end{equation}
The above structure constants suggest that we should perform the
following identifications in order to match the heterotic and M-theory
sides 
\begin{equation}
  \label{fluxid}
  f = m \; , \quad  W^a = \tfrac1{\sqrt 2} m_a \; , \quad M^a = -
  \tfrac1{\sqrt 2} \tilde m_a \; , \quad S_{ab} = m_{ab} \; .
\end{equation}
Finally the field strengths of the gauge fields can be written
explicitly using the structure constants above and we find
\begin{eqnarray}
  \label{fshet}
  F^0 & = & d A^0 \; \nn \\
  F^1 & = & d A^1 + f A^2 \wedge A^3 + M^a A^2 \wedge A^a + W^a A^3
  \wedge A^a + \tfrac12 S_{a}{}^b A^a \wedge A^b \; \nn \\
  F^2 & = & d A^2 + f A^0 \wedge A^2 - W^a A^0 \wedge A^a \; \\
  F^3 & = & d A^3 - f A^0 \wedge A^3 - M^a A^0 \wedge A^a \; \nn \\
  F^a & = & d A^a + M^a A^0 \wedge A^2 + W^a A^0 \wedge A^3 + S_b{}^a
  A^0 \wedge A^b \; \nn
\end{eqnarray}
Comparing with \eqref{fsiia} we see that we have to perform the
following identifications
\begin{equation}
  \label{gfid}
  A^0_a \leftrightarrow A^0_h \; , \qquad \tilde A^1_a \leftrightarrow
  A^1_h \; , \qquad A^2_a  \leftrightarrow - A^2_h \; , \qquad A^3_a
  \leftrightarrow A^3_h \; , \qquad A^a_a \leftrightarrow 
  \tfrac1{\sqrt 2} A^a_h \; ,
\end{equation}
where the subscript $a$ and $h$ refer to the type IIA and heterotic
pictures respectively.

In order to fully establish the above identifications we should also
check the gaugings which are produced in the case of heterotic
compactifications and compare them with the ones obtained in
M-theory. 

The scalars which are obtained from heterotic string compactification
on d-dimensional tori are arranged in $SO(d,d+16)$ matrices
\cite{MS,KM}. For the case at hand, since we split the
compactification on $T^2$ into two circle compactifications, we have
to define two such matrices. For the first $S^1$ compactification --
which we denote by the index 9 -- we find from \eqref{Tmod}
\begin{equation}
  \label{M}
  \cM_{NP} = \left(
    \begin{array}{ccc}
      g_{99} (1+ \tfrac12 g_{99}^{-1} A_9^b A_9^b)^2 & - \tfrac12
      g_{99}^{-1} A_9^b A_9^b  & A_9^a (1+ \tfrac12 g_{99}^{-1} A_9^b
      A_9^b) \\[1mm]
      - \tfrac12 g_{99}^{-1} A_9^b A_9^b & g_{99}^{-1} & - g_{99}^{-1}
      A_9^a \\[1mm]
      A_9^c (1+ \tfrac12 g_{99}^{-1} A_9^b A_9^b) & - g_{99}^{-1}
      A_9^c & \delta_{ac} + g_{99}^{-1} A_9^c A_9^a \\
    \end{array}
    \right)
\end{equation}
Here $g_{99}$ denotes the ten-dimensional metric with both legs on the
circle and $A_9^a$ denote the ten-dimensional gauge fields on
$S^1$. After this first circle compactification there will be two
additional gauge fields -- which we denote by $V^3$ and $V^4$ -- which
come from the metric and from the $B$-field respectively. The other
gauge fields which descent directly from the ten-dimensional ones we
shall denote by $V^a$. In terms of the ten-dimensional quantities they
are defined as 
\begin{eqnarray}
  \label{V8d}
  V^3_{\hat \mu} & = & g_{99}^{-1} g_{\hat \mu 9} \; , \\
  V^4_{\hat \mu} & = & B_{\hat \mu 9} + \tfrac12 A_9^a V_{\hat \mu}^a
  \; , \\
  V^a_{\hat \mu} & = & A_{\hat \mu}^a - A_9^a g_{99}^{-1} g_{\hat \mu
    9} = A^a_{\hat \mu} - A_9^a V^1_{\hat \mu} \; ,
\end{eqnarray}
where $\hat \mu$ denotes the space-time index in nine
dimensions. We shall collectively denote these vector fields as $V^N,
N= 3,4, \ldots, n_V+1$ where the index $N$ is in the fundamental
representation of the isometry group $SO(1,n_v-2)$. In 
the second circle compactification -- which we denote by the index 8
-- there will appear additional scalars coming from the
nine-dimensional vectors, $V_8^N$, and again from the metric with both
legs on the circle, $g_{88}$. Together with the scalars above they can
be assembled into a $SO(2,n_v-1)$ matrix
\begin{equation}
  \label{tM}
  \tilde \cM_{IJ} = \left(
    \begin{array}{ccc}
      g_{88} + \cM_{NP} V_8^N V_8^P + g_{88}^{-1} C^2 & g_{88} C &
      g_{88}^{-1} C L_{NR} V^R_8 + \cM_{NR} V^R_8 \\[1mm]
      g_{88} C & g_{88}^{-1} & - g_{88}^{-1} L_{NR} V^R_8 \\[1mm]
      g_{88}^{-1} C L_{PR} V^R_8 + \cM_{PR} V^R_8 & - g_{88}^{-1}
      L_{PR} V^R_8  & \cM_{NP} + g_{88}^{-1} L_{NR} L_{PQ} V^R_8 V^Q_8 \\
    \end{array}
    \right) \; ,
\end{equation}
where by $C$ we defined $C=\tfrac12 V^N_8 V^P_8 L_{NP}$ and $L$ was
defined in \eqref{L}. The isometries which are gauged in the last step
of the compactification can be read from the covariant derivatives
which are written generically as
\begin{equation}
  \label{cdM}
  D_\mu \tilde \cM_{IJ} = \partial_\mu \tilde \cM_{IJ} + f_{IK}^L
  \tilde \cM_{LJ} A^K_\mu + f_{JK}^L \tilde \cM_{IL} A^K_\mu \; ,
\end{equation}
where the non-vanishing structure constants are given in
\eqref{schet}. The above formula is nothing but the standard result
for $N=4$ gauged supergravities. The $K3$ part of the
compactification does not modify this structure. At most, some of the
gauge fields will be broken and this amounts to cut the corresponding
lines and columns from the scalar matrix above but without modifying
its structure. Thus, the result above is not so mysterious and only
descents form the $N=4$ theory which appears in tori
compactifications. The only non-trivial point so far is the assignment
of the structure constants \eqref{schet} which has been found in \cite{RES}.

The final result is nevertheless a $N=2$ gauged supergravity and in
order to be able to compare to the type IIA/M-theory side we need to
rewrite the above results in a $N=2$ language.
Using the definitions of the correct $N=2$ complex scalar fields
in four dimensions in terms of the matrix $\tilde \cM_{IJ}$ which are
given in \eqref{Tmod} it is just a matter of straightforward
algebraic manipulations to derive the form of the covariant
derivatives on these $N=2$ fields. The explicit calculation is done in
the appendix and in the following we present the final result. Setting
all the parameters in \eqref{twist2} to zero and inserting in
\eqref{cdhet2} we find 
\begin{eqnarray}
  \label{cdhet}
  D_\mu u & = &   \partial_\mu  u + f u A^0_\mu + \sqrt 2 W^a n^a A^0_\mu
  + fA^2_\mu - W^b A^b_\mu  \; , \nn \\
  D_\mu t & = & \partial_\mu t - ftA^0_\mu - \sqrt 2 M^a n^a A^0_\mu + f
  A^3_\mu + M^a A^a_\mu \; , \\
  D_\mu n^a & = & \partial_\mu n^a - \frac{M_a}{\sqrt 2} u A^0_\mu +
  \frac{W_a}{\sqrt 2} t A^0_\mu - S_a{}^b n^b A^0_\mu - \tfrac1{\sqrt
    2} \left( M_a A^2_\mu + W_a A^3_\mu - S_a{}^b A^b_\mu \right) \nn
\end{eqnarray}

It is now clear that identifying the type IIA scalar fields $x^1$, $x^2$,
$x^3$ and $x^a$ with the dilaton, $u$, $t$ and $n^a$ respectively,
and using \eqref{fluxid} and \eqref{gfid} that the two low energy
action precisely agree.

In \cite{ABLM} it was shown that the heterotic gauge field fluxes on
$T^2$ are mapped into the twist parameters $\tilde m_a$ on the
M-theory side. Here we have extended that analysis and found the
correspondent of the other twist parameters on the heterotic side. We
see that as anticipated in \cite{ABLM} the non-geometric fluxes $W^a$
which correspond to the $T$-dual of the usual fluxes $M^a$ as well as
$S_{ab}$ which correspond to twistings of the Cartan torus, find a
geometric realisation on the M-theory side as the twist parameters
$m_a$ and $m_{ab}$. 

\subsection{Generalisations: R-fluxes vs. F-theory} 

There is a certain generlisation on the heterotic side of the duality
which amounts to allow twists which would correspond to T-dualities
along non-isometric directions \cite{DH,RES}. The fluxes introduced
in this way are known as R-fluxes \cite{STW} and describe backgrounds
which do not admit geometric interpretations even locally. Formally
form the 4d perspective this would mean to introduce another twist
matrix $\tilde N$ which commutes with the initial one $N$. Let us
parameterise this new twist matrix $\tilde N$ like in \eqref{twist1}
\begin{equation}
  \label{twist2}
  \tilde N_N{}^P = \left(
  \begin{array}{ccc}
    q & 0 & U^b \\
    0 & -q & V^b \\
    -V_a & -U_a & G_a{}^b \\
  \end{array}
  \right) \; .
\end{equation}
The additional structure constants which are obtained on top of the
ones in \eqref{schet} are
\begin{equation}
  \label{scrhet}
  f_{1N}^P = \tilde N_N{}^P \; , \quad f_{NP}^0 = \tilde N_{NP} \; .
\end{equation}
The fact that the matrices $N$ and $\tilde N$ commute precisely
ensures that the structure constants in \eqref{schet} and
\eqref{scrhet} satisfy the Jacobi identity. Given the structure
constants above the $N=2$ gauged supergravity in four dimensions is
in principle fully specified as we have seen in the subsection
above. The isometries which are gauged can be 
derived from the general formula \eqref{cdM}. Following the
calculations in the appendix one finds
\begin{eqnarray}
  \label{cdhet2}
  D_\mu u  & = & \partial_\mu u + (fu + \sqrt 2 W^a n^a) A^1_\mu +
  (qu + \sqrt 2 V^a n^a) A^2_\mu  +(f - q n^a n^a - \sqrt 2U^a n^a
  u)  A^3_\mu \nn \\
  &&  - (q u^2 + \sqrt 2 V^a n^a u)A^4_\mu  
  + (V^b n^a n^a - U^b u^2 - \sqrt 2 G_b{}^a n^a u - W^b) A^b_\mu  \\[.3cm]
  D_\mu t  & = & \partial_\mu t - (ft + \sqrt 2 M^a n^a) A^1_\mu -
  (qt + \sqrt 2 U^a n^a) A^2_\mu  - (q t^2 + \sqrt 2U^a n^a t)
  A^3_\mu \nn \\ 
  &&  + (f  -q n^a n^a - \sqrt 2 V^a n^a t)A^4_\mu  
  - (U^b n^a n^a - V^b t^2 + \sqrt 2 G_b{}^a n^a t - M^b) A^b_\mu \\[.3cm]
  D_\mu n^a  & = & \partial_\mu n^a + \tfrac1{\sqrt 2} (-M^a u + W^a
  t - \sqrt 2 S_{ab} n^b) A^1_\mu + \tfrac1{\sqrt 2} (-U^a u + V^a
  t - \sqrt 2 G_{ab} n^b) A^2_\mu \nn \\
  & & + \left[-qt n^a - \sqrt 2 U^b n^b n^a - \tfrac1{\sqrt 2} U^a (ut
    - n^b n^b)  - \tfrac1{\sqrt 2} M^a \right ]A^3_\mu  \nn \\
  &  &+ \left[-qu n^a - \sqrt 2 V^b n^b n^a - \tfrac1{\sqrt 2} ^a (ut -
    n^b n^b) - \tfrac1{\sqrt 2} W^a \right ]A^4_\mu  \\
  & & + \left[(-U^b u + V^b t - \sqrt 2 G_{bc} n^c )n^a -
    \tfrac1{\sqrt 2} G_{ba} (ut - n^c n^c) + \tfrac1{\sqrt 2}
    S_{ab} \right] A^b_\mu \nn
\end{eqnarray}
Furthermore, the potential
can be computed from the $N=2$ formalism. Due to the complicated
gaugings above, the form of the potential in terms of the scalar
fields $u$, $t$ and $n^a$ is very involved and we shall
not present it here. 

Note that the potential can be computed only be relying on the $N=2$
structure of the resulting theory and can not be derived directly
from the compactification. In the heterotic case not even the source
of this potential is known and therefore being able to derive the
potential by other means may shed light on how to directly compute the
potential in such non-geometric compactifications.

Now we would like to ask whether the picture above has any sort of
type II dual. As we can see from the gaugings above, the second twist
matrix gauges isometries with respect to the gauge field $A^1$. In
type IIA we saw that such gaugings appear provided the gauge field can
be interpreted as the KK vector in a Scherk-Schwarz
compactification. Therefore we would need a second circle in the IIA
compactification. This naturally makes us consider F-theory
compactifications on $CY_3 \times T^2$. Indeed given the heterotic
F-theory duality in six dimensions we can further twist the Calabi-Yau
manifold  over the full $T^2$ precisely in the same way as we did in
the case of M-theory compactifications. Concretely, denoting the
the $T^2$ directions by $z^{1,2}$ we postulate the following
differential relations for the full 8d manifold on which we compactify
F-theory
\begin{equation}
  \label{doF}
  d \omega_i = \M_i^j \omega_j \wedge dz^1 + \tilde \M_i^j \omega_j
  \wedge dz^2 \; .
\end{equation}
Clearly, in order to assure that the exterior derivative squares to
zero the matrices $\M$ and $\tilde \M$ have to commute. Moreover the
constraint \eqref{const}  has to be satisfied by both matrices $\M$ and
$\tilde \M$. Parameterising $\tilde \M$ in the same way as we did for
the matrix $\M$, it is tempting to claim that the parameters in $\tilde
\M$ precisely correspond to the parameters from $\tilde N$. In order to
show that this picture is precisely dual to the heterotic
compactification with R-fluxes we would still need to derive the
gaugings and the structure constants of the gauge group on the
F-theory side and compare to \eqref{cdhet2} and \eqref{scrhet}. This
is not possible in general and one needs to first specify a Calabi--Yau
manifold in order to obtain a precise result from F-theory
compactifications. Therefore it seems that in general one can not test
in detail the above conjecture. There are nevertheless qualitative
analyses which point towards the fact that this duality conjecture is
right. For example, in six dimensions, after compactifying F-theory on
an elliptically fibered Calabi--Yau manifold there will be a number of
tensor multiplets which are related to the $(1,1)$ forms on the base 
of the fibration. Upon the compactification to four dimensions these
forms are supposed to satisfy the relation \eqref{doF}  and this will
imply that the corresponding tensor fields will pick up a mass in four
dimensions. In massless compactifications a tensor multiplet in six
dimensions descents to a vector multiplet in four dimensions, but in
this case we will be 
dealing with a vector-tensor multiplet where the tensor field picks up
a mass in a Stuckleberg mechanism. On the heterotic side we see no
sign of massive tensors as we were able to write the gaugings of the
$N=2$ supergravity only in terms of scalar fields. On the other hand
we have to recall that after the compactification we end up in a
different symplectic frame, and in order to be able to find agreement
between the theories we need to perform an electric-magnetic
duality. On the heterotic side this amounts to trade the gauge field
$A^1$ for its magnetic dual. For the gaugings produced by the R-fluxes
\eqref{cdhet2} we see that this gauge field appears non-trivially. In
order to perform 
the electric-magnetic duality we need to dualise first the charged
scalars to tensor fields with Green-Schwarz couplings. Then,  
by the electric-magnetic rotation these tensor fields will become
massive and it will no longer be possible to 
dualise them back into scalars. Therefore we see that in the correct
symplectic frame  we also end up with massive tensors on the heterotic
side as it was the case for the F-theory compactification.

Finally, on the F-theory side it may be possible to compute the
potential directly from the compactification. The potential will
generically contain a piece which has a geometric origin
-- coming from integrating the Ricci scalar over the eight-dimensional
manifold -- and a piece due to the four-form flux which is sourced by
the non-closure of the $(1,1)$ forms on the full eight-dimensional
space. However it is not clear if the potential can be computed in
closed form and we leave this for future research.

\subsection{Charged dilaton from fluxes}

Before closing this section we wish to make a few comments on a case
we have discarded so far. In section \ref{MS3} we have discussed
M-theory compactifications on manifolds with $SU(3)$ structure. For
the purposes of the duality with heterotic compactifications we have
chosen the parameters $m_2$ and $m_3$ in \eqref{indpar} to sum up to
zero. However, strictly from the M-theory side there is nothing to
force such a constraint upon us. After all, a twist which has $m_2+m_3
\ne 0$ is valid at least in the supergravity limit. From
\eqref{deppar} we see that the gaugings will be more complicated and
the scalar $x^1$ will also be charged. However, nothing exceptional
happens as the gaugings \eqref{kvA} were derived in general
irrespective of the solution to the constraint \eqref{const}. It is
therefore legitimate to ask what would such a choice of parameters
correspond to on the heterotic side. In fact we have seen
what is the effect in the low energy effective action if these
parameters satisfy the relation $m_2 + m_3=0$ and therefore we would
have to ask what happens in the orthogonal case, ie $m_2 - m_3 =0$.
On the heterotic side, the situation is not so simple. The M-theory field
$x^1$ corresponds to the dilaton on the heterotic side and a charged
dilaton is not common at all in heterotic string compactifications. In
fact, there are even strong no-go theorems which impose strong
constraints on the way the dilaton appears in the compactified theory.

The twist we considered in the M-theory case, $m_2 -m_3 =0$ has the
effect on the heterotic side that it takes the dilaton into minus
itself. Such a transformation is not a duality within heterotic string
theory but it is a duality of string theory, S-duality, which takes
heterotic into type I string. Implementing this duality as an allowed
twist in the compactification of the heterotic string
may lead to the same outcome as in the original
M-theory picture, namely a dialton which is charged under the
four-dimensional gauge group. Now the main question that has to be
answered is how this can be reconciled with the no-go theorems
mentioned above. These no-go theorems are based on the holomorphy
arguments and on the axionic shift symmetry of the scalar
super-partner of the dialton. This axionic field is in fact the
Poincare dual of the four-dimensional $B$-field and therefore one
expects that this shift symmetry is always present making the argument
in favor of the no-go theorem water tight. On the other hand, what
can happen in flux compactifications is that the $B$-field becomes
massive and its dualisation to an axion is no longer possible. This is
what we expect that happens in this case so that the argument for the
no-go theorem is invalidated. This can be seen easily from the
M-theory side as we shall explain in the following. 

Let us suppose that the parameters $m_a$, $\tilde m_a$ and $m_a^b$ in
\eqref{indpar} vanish and the only twist parameters which are non-zero
are $m_2$ and $m_3$ which we choose equal, ie $m_2 = m_3 = \tilde m$.
Using \eqref{deppar} and \eqref{kvA} we find the following covariant
derivatives for the scalars in the vector multiplets
\begin{eqnarray}
  D x^1 & = & d x^1 + 2 \tilde m A^1 \; \nn \\
  D x^2 & = & d x^1 - \tilde m A^2 \;  \\
  D x^3 & = & d x^1 - \tilde m A^3 \; \nn \\
  D x^a & = & d x^1 - \tilde m A^a \; \nn 
\end{eqnarray}
Now recall that in order to obtain the heterotic picture we have to
perform an electric-magnetic duality which exchanges the gauge field
$A^1$ with its magnetic dual. Since this gauge field appears
explicitly in the covariant derivative of the field $x^1$ (more
precisely only in the real part of the field), in order to perform the
electric-magnetic duality one has to promote the real part of the
field $x^1$ to a tensor field which will become massive by a
Stuckleberg mechanism. Therefore on the heterotic
side the above proposed ``S-fold'' compactification necessarily gives
rise to a massive B-field in four dimensions.

\section{Hyper-multiplet gaugings}
\label{hmg}

So far we focused on the duality in the vector multiplet sector. Now
we want to address the same question at the level of the
hyper-multiplet sector. Therefore we shall be interested in fluxes
which produce gaugings in the hyper-multiplet sector. Such fluxes in
the type IIA setting are more common than the ones discussed before
which produce gaugings in the vector multiplet sector. Still, the
lesson from the previous section will be applicable in this case and
we shall again be lead to consider M-theory or even F-theory
compactifications. We should nevertheless make it clear that the
map in the hyper-multiplet sector can not be well defined generically,
but one has to specify precisely the two backgrounds on each side of
the duality and therefore the arguments will be less accurate than in
the previous section.

\subsection{Heterotic compactifications with fluxes on K3 and their
  type IIA dual}
\label{HfK3}

In this subsection we review the known facts about the heterotic type IIA
duality with fluxes that gauge symmetries in the hyper-multiplet sector. 
Let us consider for the moment only the effects produced by the
ordinary fluxes. 
In the heterotic case such fluxes are the ones which can be turned on
inside $K3$. We denote these fluxes as
\begin{equation}
  \label{K3flux}
  F^I = m^A_I \omega_A  \; .
\end{equation}
It turns out, that in this case, the directions which are
gauged are Peccei-Quinn isometries related to the $B$-field with both
legs on $K3$, or in other words, the scalars which become charged are
the ones defined in \eqref{bexp}. We find the covariant derivatives
for these fields to be \cite{LM1} 
\begin{equation}
  \label{db}
  D b^A = d b^A - m^A_I A^I \; .
\end{equation}
Along with these gaugings a potential is generated which is in
agreement with $N=2$ supergravity as described in appendix
\ref{a1}. The action has the same form as in \eqref{s4het} with
ordinary derivatives replaced by the covariant derivatives listed
above and with a potential term added.

Let us now turn to the type IIA side. In this case there are both RR
and NS-NS fluxes which lead to gaugings in the hyper-multiplet
sector and we parameterise them as
\begin{equation}
  \label{iiaflux}
  H= q^A_0 \alpha_A - p_{0A} \beta^A \; , \qquad F_2 = m^i \omega_i \; ,
  \qquad F_4 = e_i \tilde \omega^i \; .
\end{equation}
The effect of the NS-NS fluxes $p^A$ and $q_A$ is to gauge the shift
isometries of the RR scalars $\xi^A$ and $\tilde \xi_A$ as, \cite{LM2},
\begin{equation}
  \label{dxiH}
  D \xi^A = d \xi^A - p^A A^0 \; , \qquad D \tilde \xi_A = d \tilde
  \xi_A - q_A A^0 \; .
\end{equation}
For the RR fluxes, if both $e_i$ and $m^i$ are present, then the $B$-field
is massive in four dimensions. This situation is difficult to obtain
in the heterotic picture and therefore we shall not discuss it in the
following. Hence we shall suppose that only the fluxes $e_i$ are
non-vanishing.\footnote{Note that choosing the fluxes $e_i$ is just
  for convenience as they appear on the same footing as the fluxes
  $m^i$. The two appear as electric and magnetic charges and one can
  switch between them by an appropriate electric-magnetic duality.}
The net effect of turning on such fluxes is the
presence in the four-dimensional theory of a Green-Schwarz coupling of
the type $e_i F^i \wedge B$, which upon dualisation of the $B$-field
to the universal axion $a$ shows up in the covariant derivative as
\begin{equation}
  \label{da}
  Da = da + e_i A^i \; .
\end{equation}
So far we have discussed only ordinary fluxes for the p-form field
strengths which we can turn on in both heterotic and type IIA
picture. At this stage it is far form obvious how the duality relation
is supposed to work. We shall discuss in the following various
generalisations which introduce geometric and non-geometric fluxes
and make in this way the situation more symmetric between the two
sides. Note however, that a subset of the fluxes above
should be mapped into one another as observed in \cite{CKKL}. Indeed,
since on the heterotic side the $K3$ should be elliptically fibered,
the $B$ field through the $\mathbf{P^1}$ basis of the fibration should
correspond to the universal axion, $a$, on the type IIA side. This
should make it clear that the fluxes for the gauge fields through this
cycle on the heterotic side, should precisely correspond to the
fluxes $e_i$ in \eqref{da}. The rest of the
fluxes in \eqref{K3flux} were then observed to be related to type IIA
compactifications on manifolds with $SU(3)$ structure \cite{LM3}.

Let us define the $SU(3)$ structure by deforming the harmonic forms on
the Calabi--Yau to obey \cite{GLW1} 
\begin{equation}
  \label{ghf}
  d \omega_i = q_i^A \alpha_A - p_{iA} \beta^A \; , \qquad d \alpha_A =
  - p_{iA} \tilde  \omega^i \; , \quad d \beta^A = - q_i^A \tilde \omega^i
  \; .
\end{equation}
It is not difficult to see that the fields which become charged in
this case are the RR axions $\xi^A$ and $\tilde \xi_A$. Computing $dC_3$ 
from the expansion \eqref{c3exp} and using \eqref{ghf} we immediately
find that the covariant derivatives for these fields become
\begin{equation}
  \label{dxi}
  D_\mu \xi^A = \partial_\mu \xi^A - q_I^A A^I \; , \qquad 
  D_\mu \tilde \xi_A = \partial_\mu \tilde \xi_A - p_{IA} A^I \; .
\end{equation}
In the above we have also included the $H$-fluxes from \eqref{iiaflux}
such that all the vector fields in the four-dimensional theory
participate in the gauging.
This situation resembles very much the one described in \eqref{db}
with the obvious difference that now there are two scalars in each
hyper-multiplet which are charged, compared to \eqref{db} where there
is only one charged scalar. Therefore setting half of the deformation
parameters to zero, say $q_i^A=0$, we recover precisely \eqref{db} as
it was shown in \cite{LM3}. 

The case $q_i^A\ne 0 $ does not seem to have an immediate analogue on
the heterotic side. There is a certain relaxation of the problem once
we consider manifolds with $SU(2)$ structure as we will show in the
next section, but in the most general case there will still be a
mismatch of fluxes between heterotic and type IIA sides. We expect
that this mismatch comes from the fact that we are not working with
the correct quaternionic space on the heterotic side, but only with
the sub-part which is spanned by the $K3$ moduli.

\subsection{Heterotic compactifications on manifolds with $SU(2)$
  structure and their type IIA duals} 

In this section we shall review the results of \cite{LMM} where
heterotic string compactifications on manifolds with $SU(2)$ structure
were analysed.

Let us consider that the $K3$ manifold in the heterotic
compactification is non-trivially fibered over the two-torus. In
particular we consider that the $K3$ two-forms obey
\begin{equation}
  \label{dok3}
  d \omega_A = T_{\alpha A}{}^B \omega_B \wedge dz^\alpha \; ,
\end{equation}
where $T_\alpha, ~ \alpha =1,2,$ represent the twist matrices which are
antisymmetric and commute with each other. 
In \cite{LMM} such manifolds were shown to represent almost the entire
class of manifolds with $SU(2)$ structure which have an integrable
product structure. 
The result for such a compactification is a gauged supergravity where
all the $K3$ moduli together with the fields $b_A$ are charged under
the gauge group 
\begin{equation}
  \label{cdxi}
  \begin{aligned}
    D_\mu \zeta^x_A & = \partial_\mu \zeta^x_A - T_{\alpha A}{}^B
    \zeta^x_B A_\mu^\alpha \; , \\
    D_\mu b_A & = \partial_\mu b_A - T_{\alpha A}{}^B b_B A_\mu^\alpha \; .
  \end{aligned}
\end{equation}
The vector fields $A^\alpha_\mu$ in the above formula represent the Kaluza
Klein vectors corresponding to the two circle directions $z^\alpha, ~
\alpha =1,2$.
As before, this result represents a $N=2$ gauged supergravity which
for consistency needs also a potential term. This was computed in
\cite{LMM} and was shown to precisely agree with the general formula
given in the appendix \eqref{VN2}.

Now let us turn our attention to the type IIA side. We would like to
obtain gaugings of the type \eqref{cdxi} for all the scalars in the
hyper-multiplets. We have learned that fluxes together with manifolds
with $SU(3)$ structure lead to (constant) gaugings of the shift
symmetries of the RR axions. Gaugings like the ones in \eqref{cdxi}
are not so common in type IIA 
compactifications with fluxes. However, we have encountered a similar
example in section \ref{vmg} and the way to obtain the desired 
gaugings was to lift the type IIA compactification to M-theory
compactifications on seven-dimensional manifolds with $SU(3)$
structure. Then by appropriate twists, gaugings of the type
\eqref{cdxi} can be obtained and the vector field which participates
in the gauging is the KK vector on the M-theory circle.
In the following we shall consider a similar setup, but now
twist the 3-forms around the M-theory circle \cite{LPV}, as the three forms are
the ones which mostly govern the hyper-multiplet sector in type IIA
compactifications.\footnote{M-theory compactifications on manifolds with
$SU(3)$ structure which give rise to potentials for the hyper-scalars
were first studied in \cite{MPS}.}
Recall that the three-forms on a Calabi--Yau
manifold which satisfy \eqref{norm} can be
rotated by a symplectic transformation. We shall use this symplectic
symmetry in defining the twisting. Consider the following dependence
on the M-theory coordinate
\begin{equation}
  \label{stw}
  d \left(
  \begin{array}{c}
    \alpha_A \\[.2cm]
    \beta^A
  \end{array}
  \right) = \left (
    \begin{array}{cc}
      M_A{}^B & 0 \\[2mm]
      0 & - M_B{}^A
    \end{array}
    \right) \cdot \left(
      \begin{array}{c}
        \alpha_B \\[2mm]
        \beta^B
      \end{array}
      \right) \wedge dy \; ,
\end{equation}
where the twist matrix is symplectic by construction. Here we denoted
by $y$ the circle direction in order to avoid confusion with the
complex structure moduli of the Calabi--Yau manifold. In order to
match the heterotic side we shall also consider that the matrix
$M_A{}^B$ is also antisymmetric. A more general symplectic twist does
not seem to have an immediate analogue on the heterotic side, but in
order to make a more precise statement one would need to have an
explicit map of the hyper-multiplets. While the most general case was
discussed in \cite{LPV}, in the following we limit
ourselves to the Ansatz \eqref{stw} which can be written explicitly
\begin{equation}
  \label{dab}
  d \alpha_A = M_A{}^B \alpha_B \wedge dy \; , \qquad 
  d \beta^A = - M_B{}^A \beta^B \wedge dy \; .
\end{equation}
Note that this automatically preserves the orthonormation of the forms
$\alpha$ and $\beta$ which on the full seven-dimensional manifold
reads
\begin{equation}
  \int_{7d} \alpha_A \wedge \beta^B \wedge dy = \delta_A^B \; .
\end{equation}

Now we proceed in close analogy to \cite{ABLM} and we shall find it
more convenient to work with a basis of forms which does not depend on
the additional M-theory coordinate and transfer all this dependence on
the moduli fields. 

Let us concentrate on the scalars in the hyper-multiplets. These fields
come from expanding the holomorphic $(3,0)$ form $\Omega$ and the
three-form gauge potential $C_3$ in the basis of three-forms like in
\eqref{Oexp} and \eqref{c3exp}. Gauge
invariance requires that the fields $Z^A$, $\cG_A$, $\xi^A$ and $\tilde \xi_A$
transform as
\begin{equation}
  \label{trsfh}
  \delta Z^A = - M_B{}^A Z^B \epsilon \; , \quad \delta \cG_A = M_A{}^B
  \cG_B \epsilon \; , \quad \delta \xi^A = -M_B{}^A \xi^B \epsilon \; ,
  \quad \delta \tilde \xi_A = M_A^B \tilde \xi_B \epsilon \; , 
\end{equation}
under the change $y\to y + \epsilon$.
Let us make a couple of comments here. First, note that the
transformation of $\cG_A$ above is required by gauge invariance. On the
other hand, $\cG_A$ are the derivatives of the prepotential $\cG$ with
respect to $Z^A$ and therefore one can infer its transformation from
the definition of $\cG$. Using the holomorphy of the prepotential $\cG$ one
immediately finds 
\begin{equation}
  \delta \cG_A = \cG_{AB} \delta Z^B = - \cG_{AB} M_C{}^B Z^C \epsilon \; . 
\end{equation}
Comparing with the corresponding transformation from \eqref{trsfh} we
find 
\begin{equation}
  - \cG_{AB}M_C{}^B = M_A{}^B \cG_{BC} 
\end{equation}
This means that for a given prepotential, the possible twists are
given by the solutions to the above constraint. This is precisely the
analogue of the condition \eqref{const} found in the previous section
when twisting the harmonic $(1,1)$ forms of the Calabi--Yau manifold
over the circle. For a generic
prepotential we expect no isometry of the special geometry defined by
it and therefore no matrix $M$ will satisfy the constraint. Here
however we shall consider that there are certain isometries of the
special geometry and therefore the constraint will have non-trivial
solutions. We do this assumption because ultimately we are interested
in mapping this compactification to heterotic strings on $K3 \times
T^2$ and we know that the quaternionic space of the hyper-scalars
contains the $K3$ moduli space $SO(4,20)/SO(4) \times SO(20)$ which
originates from a special geometry of the type $SU(1,1)/U(1) \times
SO(2,18)/SO(2)\times SO(18)$.

The second observation we want to make here is that not all the $Z^A$
fields are independent degrees of freedom because they are only
projective coordinates on the space of complex structure
deformations. In many of the calculations it is convenient to fix the
gauge by choosing $Z^0=1$. For this to be possible in the present
context we need that $Z^0$ does not transform under $y\to y+
\epsilon$. In the following we shall choose to fix the gauge $Z^0=1$
at the expense of setting $M^0{}_A = M^A{}_0=0$. From the perspective
of the duality with the heterotic compactifications this is not so bad
as the parameters we want to set to zero lead to gaugings of the
scalars in the universal hyper-multiplet which is special anyway 
and we do not focus on it here. 

Consequently they will have covariant derivatives of the form
\begin{eqnarray}
  \label{cdhy}
  D_\mu z^a & = & \partial_\mu z^a - M_b{}^a z^b A^0 \; ; \nn \\
  D_\mu \xi^a & = & \partial_\mu \xi^a - M_b{}^a \xi^b A^0 \; ; \\
  D_\mu \tilde \xi_a & = & \partial_\mu \tilde \xi_a + M_a{}^b \tilde
  \xi_b A^0 \; , \nn  
\end{eqnarray}
where now $a=1, \ldots , n_h$.
We see that the result above has precisely the same form as the
gauging \eqref{cdxi}. The only difference comes from the fact that in
the heterotic case there were two twist matrices $T_\alpha, ~
\alpha=1,2$, while 
in \eqref{cdhy} there is only one. A similar problem we have
encountered in the previous section where we were trying to match the
gaugings in the vector multiplet sector in heterotic and type IIA
compactifications. There we have argued that in order to restore the
duality one has to go all the way up to F-theory compactifications on
eight-dimensional manifolds which are obtained from fibering the
Calabi--Yau manifold over a $T^2$. If we apply the same logic here we
will have to introduce a second twist matrix $\tilde M$ which commutes
with the matrix $M$. The result then will precisely match the
heterotic side. So also in the hyper-multiplet sector the most general
gauging which can be obtained on the heterotic side from
compactifications on manifolds with $SU(2)$ structure can be mapped to
F-theory compactifications. In this case however, the duality only relates
geometric backgrounds and there is no non-geometric aspect involved. 

\subsection{Turning on multiple fluxes}

So far we have only turned on very specific types of fluxes at a time
and we did not analyse what happens if we try to turn on more fluxes
simultaneously.

First of all let us note that in the last case studied, a vev for the
scalars $\xi$ and
$\tilde \xi_a$ produces a term in the covariant derivatives
\eqref{cdhy} similar to the one in \eqref{dxi}. Therefore, the effect
of the gaugings in \eqref{dxi} can be simply removed by shifting the
vev for the scalars $\xi^A$ and $\tilde \xi_A$ in an appropriate
way. In this way, the geometric fluxes obtained by compactifying
M-theory on manifolds with $SU(3)$ structure are more fundamental than
the fluxes introduced in \eqref{ghf}. Since the fluxes in \eqref{dab}
and the corresponding F-theory deformations, are simply mapped to the
twists \eqref{dok3} on the heterotic side, it means that the effect of
some of the fluxes in \eqref{ghf} can be obtained on the heterotic
side by simply considering manifolds with $SU(2)$ structure and
shifting the vev of certain scalars. 
Thus, this resolves a part of the puzzle
encountered at the end of section \ref{HfK3}.

More generally the fluxes \eqref{ghf} and the twisting \eqref{dab} are
not compatible. To see this note that acting with the exterior
derivative on $\omega_i$  twice we need that 
\begin{equation}
  p_{iA} M_B{}^A =0 \; , \qquad \mathrm{and} \qquad q_i^A M_A{}^B =0 \; . 
\end{equation}
This means that the twist matrix $M$ must have zero eigenvalues or in
other words it means that some of the three-forms or some combination
thereof do not change as we go around the circle. These forms will be
precisely the ones which are allowed to appear on the right hand side
of \eqref{ghf}.

The same exclusion between fluxes can be seen also on the heterotic
side. Here the gauge field fluxes on $K3$ are in general incompatible
with the twisting discussed in this section. The reason is the Bianchi
identities the field strengths must satisfy\footnote{Note that we are
  dealing with fluxes for Abelian gauge fields}
\begin{equation}
  d F^I =0 \; .  
\end{equation}
If we try to turn on both the fluxes \eqref{K3flux} and the twisting,
the above Bianchi identity will imply 
\begin{equation}
  m^I_A T_{\alpha B}{}^A = 0
\end{equation}
which is precisely of the same form as in type IIA case and tells us
that gauge field fluxes can only be turned on along eigenvectors of
$T_i$ corresponding to zero eigenvalue.

On the other hand, fluxes which gauge isometries in the vector
multiplet sector and those gauging isometries in the hyper-multiplet
sector can coexist and their effects can be simply added up.

\subsection{Leftover fluxes}

Until now we have only discussed the fluxes which have a dual
interpretation. It is also important to review which are the
fluxes for which the dual is not known. As long as we are talking about
vector multiplet gaugings we have seen in section \ref{vmg} that for all
the fluxes there is at least a proposal for their dual. For the
hyper-multiplets gaugings the situation is not so simple and there are
several fluxes for which a dual is not known. Recall that at the
beginning of section \ref{hmg} we set the parameters  $m^i$ in
\eqref{iiaflux} to zero. These fluxes introduce magnetic gaugings and
it is not clear how to obtain something similar on the heterotic
side. The same applies to fluxes which come from compactifying type IIA
on manifolds with $SU(3) \times SU(3) $ structure. These fluxes
introduce charges which are magnetic dual to the ones in \eqref{dxi}
and do not have a known heterotic dual. Finally we have already
explained that only half of the flux parameters in \eqref{dxi} have dual
heterotic interpretation. 

On the heterotic side we can use manifolds with $SU(2)$ structure which
do not have an integrable product structure and the corresponding
fluxes do not seem to have a type IIA dual. However it is not clear
whether such manifolds with $SU(2)$ structure are meaningful from the
point of view of the heterotic type IIA duality.

\section{Conclusions}

In this paper we have studied the fluxes which can be turned on in
heterotic and type IIA compactifications to four dimensions with $N=2$
supersymmetry from the point of view of the heterotic type IIA
duality.

We distinguished two classes of fluxes: fluxes which gauge isometries
in the vector multiplet sector and fluxes which gauge isometries in
the hyper-multiplet sector. In section \ref{vmg} we extended the
results in \cite{ABLM} and showed that all the fluxes which appear in
M-theory compactifications on manifolds with $SU(3)$ structure and
which gauge isometries in the vector multiplet sector have a
correspondent in heterotic compactifications with duality twists as
discussed in \cite{DH,RES}. Such fluxes include among others
non-geometric fluxes which can be intuitively understood as fluxes for
the T-dual gauge fields and have a purely geometric origin on the
M-theory side. Furthermore we conjectured that the heterotic R-fluxes
introduced in \cite{RES} find a geometric realisation on the other
side of the duality in the framework of F-theory compactifications on
eight-dimensional manifolds with $SU(3)$ structure. For this
conjecture we have only presented a few indications including the
counting of flux degrees of freedom. A more detailed analysis is
needed especially on the F-theory side in order to be sure that the
proposed scenario is indeed the correct one. However F-theory
compactifications highly depend on the Calabi--Yau manifold used and a
general analysis is not possible. One may still use a dual
picture,\footnote{We thank Eran Palti for pointing this out.}  
like M-theory compactified to three-dimensions,\footnote{See \cite{TG}
  for recent work on this topic.} but this is beyond the scope of the
present paper. 

For the hyper-multiplet sector the duality map is not very well
specified and therefore there is a certain ambiguity in finding the
dual fluxes. Still we have been able to identify large classes of
fluxes which can be mapped from the heterotic to type IIA side. Again,
like in the vector multiplet sector, certain fluxes seem to correspond
to M and even F-theory compactifications on manifolds with $SU(3)$
structure. 

Dualities in string compactifications with fluxes have played an
important role in understanding various fluxes which can be turned on
in different situations. We have also seen it in the present paper
that by duality we are lead to consider new fluxes in the same spirit
as \cite{AACG}. From this point
of view it would be interesting to find the most general setup which
is fully invariant under this duality. 

Finally we want to comment on the practical use of the results in this
paper. Since we are dealing with $N=2$ supergravities the possibility
to apply these results to phenomenology is quite remote. We
nevertheless want to point out that even the $N=2$ analysis may be
useful at least for deriving the low energy effective actions in
$N=1$ compactifications which use similar backgrounds. Also one may
consider various projections/truncations similar to the orientifolding
in type II compactifications such that the final theory has only $N=1$
supersymmetry. Last but not least, one may check if the theories
described here exhibit spontaneous $N=2 \to N=1$ breaking
\cite{LST}. From this point o view, the only setup which may be suitable
for such an analysis is heterotic compactifications with $R$-fluxes as
described in section \ref{vmg}. Indeed as explained there, heterotic
compactification naturally take us into a symplectic frame where no
prepotential exists and moreover rotations to a basis where a
prepotential exists induce magnetic gaugings along with the existing
electric ones which is a necessary condition for a $N=2$ supergravity
to present spontaneous $N=2 \to N=1$ breaking.

\vspace{1cm}
\noindent
{\textbf{Acknowledgments}}
This work was supported in part by the National University Research
Council CNCSIS-UEFISCSU, project number PN II-RU 3/3.11.2008 and PN
II-ID 464/15.01.2009 and in part by project ''Nucleu'' PN 09 37 01 02
and PN 09 37 01 06.
The author thanks Emilian Dudas, Mariana Gra\~{n}a, Ruben Minasian and
Eran Palti for helpful discussions.

\vspace{1cm}
\appendix
\noindent{\Large\textbf{Appendix}}

\section{$N=2$ (gauged) supergravity in four dimensions}
\label{a1}

In this appendix we shall review the main features of $N=2$
supergravity in four dimensions. As the compactifications we are
dealing with fall in this class the formulae here will be applicable
to both  type IIA and heterotic pictures. We shall only be concerned
with the bosonic fields and therefore we shall largely ignore the
fermions whose interactions can be obtained by supersymmetry.

The $N=2$ supergravity multiplet contains the graviton $g_{\mu \nu}$
and a vector field, the graviphoton. Other $N=2$  multiplets which will
be of interest for us are the vector multiplets and the
hyper-multiplets. The vector multiplets contain one vector field and
one complex scalar in the adjoint of the gauge group. The
hyper-multiplets contain four scalar fields and are responsible for the
matter content of the theory.  $N=2$ supersymmetry requires that the
manifold spanned by the scalars splits into a product of special
K\"ahler manifold -- which describes the scalars in the vector
multiplets -- and a quaternionic manifold which describes the
hyper-scalars
\begin{equation}
  \label{smN2}
  \cM_{scalar} = \cM_{SK}  \times \cM_{Q} \; .
\end{equation}

For the issues discussed in this paper, the quaternionic
manifold does not play any special role. In type IIA compactifications
the quaternionic metric can be written explicitly in terms of
quantities defined on the Calabi--Yau manifold, while in the heterotic
case the metric is not known in general. 

On the special-K\"ahler manifold $\cM_{SK}$ we can introduce
projective coordinates, $X^I, I= 0, \ldots, n_v$, in terms of which
the scalars in the vector multiplet sector are given by 
\begin{equation}
  \label{eq:6}
  x^i = \frac{X^i}{X^0} \; , \quad i = 1, \ldots , n_v \; ,
\end{equation}
where $X^0$ is supposed to be non-vanishing. The geometry is then
described entirely by a holomorphic function, called prepotential,
$\cF(X^I)$, which is homogeneous of degree two in the projective
variables $X^I$. The K\"ahler potential is given in terms of the
prepotential as
\begin{equation}
  \label{KN2}
  K = - \ln{(X^I \bar \cF_I - \bar X^I \cF_I)} \; ,
\end{equation}
where $\cF_I= \partial_{X^I} \cF $ denote the derivatives of the
prepotential. Moreover, the same function $\cF$ gives the couplings of
the gauge fields in the vector multiplets
\begin{equation}
  \label{NN2}
  \cN_{IJ} = \bar \cF_{IJ} + 2i \frac{Im \cF_{IK} Im \cF_{JL} X^K X^L}{Im
    \cF_{KL} X^KX^L}
\end{equation}
where $\cF_{IJ} = \partial_{X^I} \partial_{X^J} \cF$. The imaginary part
of the above matrix describes the generalised 
coupling constants while the real part the generalised theta
angles. Altogether the bosonic part of the $N=2$ supergravity action
is given by
\begin{equation}
  \label{s4het}
   S = \int \Big[ \frac12 R * \mathbf{1} 
   - g_{i\bar\jmath} d x^i \wedge * d \bar{x}^{\bar\jmath} 
   - h_{uv} d q^u \wedge *d q^v
  + \frac{1}{4}\, Im \cN_{IJ} F^I \wedge * F^J  
  + \frac{1}{4} \, Re \cN_{IJ} F^I \wedge F^J \Big]  \ .
\end{equation}
Any (global) symmetry of this action must necessarily be an isometry of the
scalar manifold \eqref{smN2}. Some of these symmetries can be made
local (gauged) and in this case the partial derivatives in the
kinetic terms for the scalars are replaced by appropriate covariant
derivatives 
\begin{equation}
  \label{cdN2}
  d x^i \to D x^i = d x^i - k^i_I A^I \; ; \qquad d q^u \to D q^u = d
  q^u - k^u_I A^I \; ,
\end{equation}
where $k_I^u$ and $k_I^i$ are the components of the Killing vectors 
which give the directions in the scalar space which are
gauged. The holomorphic Killing vectors $k_I^i$ can be obtained as
derivatives of a holomorphic prepotential $\cP_I$, while $k_I^u$ can be
obtained as covariant derivatives (on the quaternionic space of
hyper-multiplets) of a triplet of prepotentials $\cP_I^x$.
Since they do not play any role in the paper we shall not
discuss them in the following and refer the interested reader to the
existing literature \cite{N2rev}. Finally, $N=2$ supersymmetry
requires the presence of a scalar potential potential in connection
with the gaugings above, which in terms of the prepotentials defined
above is given by
\begin{equation}
  \label{VN2}
  V = e^K X^I \bar X^J \left(g_{\bar \imath j} k^{\bar \imath}_I k^j_J
    + 4 h_{uv} k^u_I k^v_J \right) - \left[ \tfrac12 (Im \cN)^{-1 \;
      IJ} + 4e^KX^I X^J \right] P_I^x P_J^x \; .
\end{equation}

In the case when the gauge group is non-Abelian, the fields
$X^I$ are charged as they transform in the adjoint representation of
the gauge group along with the vector fields. If the prepotential
defining the special K\"ahler geometry, $\cF$, is
invariant under the gauge transformations, then the $N=2$ action is
obviously invariant under gauge transformations. However it is also
possible that the prepotential is not invariant and one can still 
define an invariant action. This is the case when the prepotential
changes under gauge transformations by a term 
\begin{equation}
  \label{delF}
  \delta \cF = \Lambda^I C_{IJK} X^J X^K \; ,
\end{equation}
where $\Lambda^I$ denote the gauge transformation parameters and
$C_{IJK}$ are real constants. For the transformation above, the
K\"aher potential changes by the real part of a holomorphic function
(thus leaving the K\"ahler metric invariant) while 
kinetic terms for the gauge fields are left invariant. However, the
generalised theta terms do change and in order to reestablish the
gauge invariance of the action one has to add the following term
\begin{equation}
  \label{CSN2}
  \frac13 \int C_{IJK} A^I \wedge A^J \wedge \left(d A^K - \frac38
    f_{LM}^K A^L \wedge A^M \right) \; .
\end{equation}

In the end let us discuss the electric magnetic duality of $N=2$
supergravities which is of central importance to the work presented
here. This duality does not represent an invariance of the action but
rather a symmetry of the equations of motion together with the Bianchi
identities. Let us define the magnetic fiend strengths as
\begin{equation}
  \label{mfs}
  G_I = \frac{\partial \mathcal{L}}{\partial F^I} \; ,
\end{equation}
where $\mathcal{L}$ denotes the Lagrange density of the $N=2$
supergravity theory. The system of equations of motion and Bianchi
identities (in the ungauged theory) which read
\begin{equation}
  \label{eomBI}
  d G_I = 0 \; , \qquad d F^I = 0 \; ,
\end{equation}
is invariant under symplectic rotations
\begin{equation}
  \label{srFG}
  \left(
    \begin{array}[h]{c}
      F^I \\
      G_I\\
    \end{array}
  \right) \to
  \left( 
    \begin{array}[h]{cc}
      U & Z \\
      W & V \\
    \end{array}
  \right)
  \left(
    \begin{array}[h]{c}
      F^I \\
      G_I\\
    \end{array}
  \right) \; .
\end{equation}
where $U$, $V$, $W$ and $Z$ are constant, real matrices which obey
\begin{equation}
  \begin{aligned}
    U^T V &- W^T Z  = V^T U - Z^T W = \mathbf{1} \; , \\
    U^T W & = W^T U \; , \qquad Z^T V = V^T Z \; .
  \end{aligned}
\end{equation}
Under this transformation $(X^I, \cF_I)$ form a symplectic vector
which transforms precisely as the vector $(F^I, G_I)$. Finally, the
gauge coupling matrix $\cN$ transforms as
\begin{equation}
  \label{spN2}
  \cN \to (V\cN + W) (U + Z \cN)^{-1} 
\end{equation}
It can be easily seen that in general there exist
symplectic transformations such that in the resulting frame $\cF_I$ are
no longer the derivatives of one function $\cF$ and therefore in such
frames, the prepotential does not exist. Nevertheless, the vector
$(X^I, \cF_I)$ is enough for defining the geometry and the K\"ahler
potential is again given by the formula \eqref{KN2}. Compactifications of
the heterotic string on $K3 \times T^2$ manifolds naturally leads to
special geometries in a frame where no prepotential exists.

In the gauged theory as presented above, the symplectic symmetry is
broken by choosing the ``electric'' vector fields which participate in
the gauging.\footnote{There exist a formalism -- aka the embedding tensor
formalism \cite{dWST}-- in which the $N=2$ gauged supergravity is written in an
explicit symplectic covariant way. For this one introduces magnetic
gaugings and allows the electric and magnetic charges to be rotated
into one another by the symplectic rotations.} The fact that in the
heterotic case one ends up in a symplectic frame where no prepotential
exists is crucial for explaining the duality with type IIA
compactifications.

\section{The  vector multiplet sector in the heterotic compactifications} 

\label{appB}

Much of the structure of the vector multiplet sector in heterotic
compactifications on $K3 \times T^2$ comes from the compactification
on $T^2$. In general $T^d$ compactifications have 16 supercharges
(hence $N=4$ in four dimensions) and therefore the vector multiplet
sector inherits much of the structure of $N=4$ theories. In particular
the torus moduli together with the Wilson lines parameterise a
$O(d,d+16)$ matrix as follows \cite{MS}
\begin{equation}
  \label{Tmod}
  \cM_{IJ} = \left(
    \begin{array}{ccc}
      g + C^T g^{-1} C + A A^T& - C^T g^{-1} & C^T g^{-1} A + A \\
      - g^{-1} C & g^{-1} & - g^{-1} A \\
      A^T g^{-1} C + A & -A^T g^{-1} & \mathbf{1}_{16 \times 16} + A^T
      g^{-1} A
    \end{array}
    \right)
\end{equation}
In the above we have used matrix notation (with matrix multiplication
assumed) and the various quantities have the following meaning: $g$ is
a $d \times d$ matrix representing the metric on the $d$-dimensional
torus, $A$ is a $d \times 16$ matrix made of the gauge fields $A^a, a
=1, \ldots, 16$ with legs on $T^d$ and $C$ is a $d \times d$ matrix
which is given by $C= B + \tfrac12 A A^T$, where $B$ denotes the
$B$-field on $T^d$.

We are obviously interested in the case $d=2$ which means that the
matrix above is a $20 \times 20$ matrix. We also have to keep in mind 
that the $K3$ part of the compactification
may influence the above by the fact that the gauge bundle which we
need to turn on, may break some of the original gauge symmetry and
therefore the dimension of the scalar space may not be 36 as
it should have been in this case, but can be smaller.\footnote{In
  principle, some of the KK gauge fields which appear in the $T^2$
  compactification may be broken by the gauge bundle on $K3$. We shall
  however consider these gauge fields survive the compactification as
  they constitute the most interesting sector when fluxes are turned
  on.}  Consequently the matrix $\cM$ above may 
be smaller and we shall work with a moduli matrix
parameterising a $O(2,n_v-1)$ group element. As in the $N=4$ theory,
the kinetic term for the moduli is given as the standard kinetic term
on a group, namely $tr(\partial_\mu \cM \partial^\mu \cM)$. In the
$N=2$ theory, the scalar manifold spanned by the scalars in the vector
multiplets is a special K\"ahler manifold and it is meaningful to
write the kinetic term in the appropriate way. This is done by
defining complex coordinates on the space as
\begin{eqnarray}
  \label{complex}
  A^a_9 & = & \sqrt 2 \frac{n^a- \bar n^a}{u - \bar u}  \; ; \qquad
  A^a_8 = \sqrt 2  \frac{\bar u n^a - u \bar n^a}{u - \bar u}\nn \\ 
  B_{98} & = &\tfrac12 \left[(t + \bar t) - \frac{(n^a + \bar n^a)(n^a
      - \bar n^a)}{u -\bar u} \right]  \\
  \sqrt g & = & - \tfrac{i}{2} \left[(t - \bar t) - \frac{(n^a - \bar
      n^a)(n^a - \bar n^a)}{u - \bar u} \right] \nn \\
  g_{99} & = &\frac{2i}{u - \bar u} \sqrt g  \; ; \qquad g_{89} = i
  \frac{u + \bar u}{u - \bar u} \sqrt g \nn
\end{eqnarray}
Finally there is the dialton field, $\phi$, together with its partner,
the axion, $a$, which is Poincare dual to the $B$-field in four
dimensions which combine as
\begin{equation}
  s= a - i e^\phi \; .
\end{equation}

The special geometry data is now given as
\begin{eqnarray}
  \label{XFhet}
  X^0 = 1 \; ; \quad X^1 = ut - n^a n^a \; ; \quad X^2 = -u \; ; \quad
  X^3 = t \; ; \quad X^a = \sqrt 2 n^a \; ; \\
  \cF_0 = s(ut -n^2 ) \; ; \quad \cF_1 = s  \; ; \quad \cF_2 = st \; ;
  \quad \cF_3 = - su \; ; \quad \cF_a = \sqrt 2 s n^a\; ;
  \nn 
\end{eqnarray}
Note that the symplectic frame above is such that $\cF_I$ are not
derivatives of a prepotential and therefore in this basis no
prepotential exists. By a symplectic transformation ($X^1 \to \tilde
X^1 = \cF_1$ and $\cF_1 \to \tilde \cF_1 = - X^1$) we reach a basis
where a prepotential exists. This is precisely the type IIA basis and
the prepotential is given by \eqref{prepot} with intersection numbers
\eqref{intno}. 

To end this section, note that because in the natural heterotic
symplectic basis there is no prepotential, the gauge coupling matrix
can not be computed as in \eqref{NN2}. One can use the more
complicated formalism of \cite{N2rev} or use a detour to go first to a
symplectic frame where a prepotential exists, compute the gauge
coupling matrix there and then perform the inverse rotation to the
original symplectic frame and use formulae \eqref{spN2} to transform
the gauge coupling matrix. One can also obtain the gauge coupling
matrix directly from the compactification and the result is given by
\begin{equation}
  \label{Nhet}
  Im \cN_{IJ} = \frac{s - \bar s}{2i} \cM_{IJ} \; , \qquad Re \cN_{IJ}
  = - \frac{s + \bar s}{2} \eta_{IJ} \; ,
\end{equation}
for the matrix $\cM_{IJ}$ defined in \eqref{Tmod} for the special case
of $T^2$ compactifications.

\subsection{Gaugings in the vector multiplet sector}

The purpose of this appendix is to show the calculations which lead to
the gaugings \eqref{cdhet2} and implicitly to \eqref{cdhet} in the
particular case when there are no R-fluxes. The calculation is
straightforward and one starts form the general form for the covariant
derivative \eqref{cdM} and substitutes the matrices \eqref{tM} and
\eqref{M} using the definitions for the fields \eqref{complex}. In
order to ease the calculation it is worth noting that the scalars
$V^M_8$ have a simple form when expressed in terms of the fields $t$,
$u$ and $n^a$. In particular we find
\begin{equation}
  \label{V8}
  V_8^3 = \frac{u + \bar u}{2} \; ; \qquad  V_8^4 = - \frac{t + \bar
    t}{2} \; ; \qquad V_8^a = - \frac{n^a + \bar n^a}{\sqrt 2} \; . 
\end{equation}

Let us now expand formula \eqref{cdM} using the matrices \eqref{tM}
and \eqref{M} with the structure constants \eqref{schet} and
\eqref{scrhet}. From \eqref{V8} we see that the covariant derivative
of the element $\tilde \cM_{2M}$ will give us valuable information
about the covariant derivatives of the fields $u$, $t$ and
$n^a$. Actually we shall argue that using properties like the
holomorphy of Killing vectors, the evaluation of this covariant
derivative will be enough to determine without any ambiguity the form
of the covariant derivatives of the fields $u$, $t$ and $n^a$. In
order to evaluate the covariant derivative of the matrix element
$\tilde \cM_{2N}$ we first need to write the one of the element
$\tilde \cM_{22}= g_{88}^{-1} \equiv g^{88}$. We find
\begin{equation}
  \label{cdg88}
  D_\mu g^{88} = \partial_\mu g^{88} - 2 g^{88} \tilde N_{NP} V^P_8
  A_\mu^N \; .
\end{equation}
Note that this covariant derivative is non-trivial only if R-fluxes are
present. 

Before we compute the covariant derivative of the element $\tilde \cM_{2N}$
we should clarify one point about the notations we use. The capital
indices $I$, $J$, etc. in formula \eqref{cdM} run over all the vector
fields in the theory while the indices $N$, $P$, etc. run over two
less vector fields (ie the ones which appear in the last step of the
compactification from five to four dimensions). We have therefore used
the notation $I= (1,2, N)$. Therefore the indices $N$, $P$, etc. are
understood to range form $3, \ldots n_v$. As an example, the first
element of the matrix $\cM_{NP}$ -- ie the element $(1,1)$ in standard
notation -- will be denoted by $\cM_{33}$.

With the above observations, from the covariant derivative of the
matrix element $\tilde \cM_{1N}$ we find
\begin{eqnarray}
  D_\mu \left(L_{NP}V^P_8 \right) & = & - \left(D_\mu g_{88} \right)
  \tilde \cM_{2N}  - g_{88} \left(D_\mu \tilde \cM_{2N} \right) \nn \\
  & = & \partial_\mu \left(L_{NP} V^P_8 \right) - N_{NQ} \left(V^Q_8
    A^1_\mu + A^Q \right) - \tilde N_{NQ} V^Q_8 A^2_\mu  \\
  & & + \left(\tilde N_{QR} V^R_8
  L_{NP}V^P_8 - g_{88} \tilde N_Q{}^R \cM_{RN} + \tfrac12 \tilde
  N_{NQ} L_{PR} V^P_8 V^R_8 \right) A^Q_\mu \nn 
\end{eqnarray}
We can now specialise for various values of the index $N=3,4,\ldots,
n_v+1$. Using the definitions \eqref{complex} in order to express the
elements of the matrix $\cM$ and the parameterisation of the twist
matrices \eqref{twist1} and \eqref{twist2} we find
\begin{eqnarray}
  D_\mu (u + \bar u) & = & \partial_\mu (u + \bar u) - \tilde N_{J4}
  (n^a n^a + \bar n^a \bar n^a) A^J_\mu  + \tilde N_{J3} (u^2 + \bar
  u^2) A^J_\mu \\
  & & - \sqrt2 \tilde N_{Ja} (n^a u + \bar n^a \bar u) A^J_\mu - 2
  N_{4J} (A^J_\mu + V^J_8 A^1_\mu) - 2 \tilde N_{4J} V^J_8 A^2_\mu \nn
  \\[.2cm]
  D_\mu (t + \bar t) & = & \partial_\mu (t + \bar t) - \tilde N_{1J}
  (n^a n^a + \bar n^a \bar n^a) A^J_\mu  - \tilde N_{J4} (t^2 + \bar
  t^2) A^J_\mu \\ 
  & &  - \sqrt2 \tilde N_{Ja} (n^a t + \bar n^a \bar t) A^J_\mu + 2
  N_{3J} (A^J_\mu + V^J_8 A^1_\mu) + 2 \tilde N_{3J} V^J_8 A^2_\mu \nn
  \\[.2cm]
  D_\mu (n^a + \bar n^a) & = & \partial_\mu (n^a + \bar n^a) + \tilde
  N_{J3} (n^a u + \bar n^a \bar u) A^J_\mu - \tilde N_{J4} (n^a t +
  \bar n^a \bar t) A^J_\mu \\
  & & - \sqrt 2 \tilde N_{Jb} (n^b n^a + \bar n^b \bar n^a) A^J_\mu -
  \tfrac1{\sqrt 2} \tilde N_{Ja} (ut + \bar u \bar t - n^b n^b - \bar
  n^b \bar n^b) A^J_\mu \nn \\
  & & + \tfrac1{\sqrt 2} \tilde N_{aJ} V^J_8 A^2_\mu + \tfrac1{\sqrt
    2} N_{aJ} (A^J_\mu + V^J_8 A^1_\mu) \nn 
\end{eqnarray}
Similarly, using other elements of the matrix $\tilde \cM$ one can
derive the covariant derivatives of the imaginary parts of the
fields. However, using the holomorphy of the Killing vectors we can
already read off from the expressions above what the covariant
derivatives of the complex fields $u$, $t$ and $n^a$ are. The only
ambiguity can come from the constant terms in Killing vectors. However,
such constant terms can not appear in the covariant derivatives of the
imaginary parts of the fields $u$, $t$, and $n^a$, as such terms
correspond to gaugings of 
shift isometries and the imaginary parts of the fields do not have
such invariances in the ungauged theory.\footnote{Note that the
  imaginary parts of the fields appear explicitly in the K\"ahler
  potential and therefore the theory does not have shift symmetries in
  these directions.}
With this one immediately sees that the relations above imply the
covariant derivatives written in \eqref{cdhet2}
As a consistency check, one can verify, after a lengthy, but
completely straightforward calculation, that the Killing vectors which
can be read from these equations satisfy the commutation relations
\begin{equation}
  \label{commut}
  [k_I, k_J] = f_{IJ}^K k_K \; ,
\end{equation}
with the structure constants $f_{IJ}^K$ defined in \eqref{schet} and
\eqref{scrhet}.

\end{document}